\documentclass[a4paper]{article}

\usepackage{INTERSPEECH2020}
\usepackage{multirow}
\usepackage{color}
\usepackage{subcaption}
\usepackage{url}
\usepackage{etoolbox}
\usepackage{pgfplotstable}
\usepackage{tikzscale}
\usepackage{tikz}
\pgfplotsset{compat=newest}
\usetikzlibrary{external,arrows,calc,positioning,shapes.geometric,math}

\newtoggle{release}
\togglefalse{release}

\iftoggle{release}{
\newcommand{\todo}[1]{}
\newcommand{\ya}[1]{}
\newcommand{\gab}[1]{}%
}{
\newcommand{\todo}[1]{\textcolor{red}{\textbf{\underline{alex}}: #1}\PackageWarning{TODO:}{#1!}}
\newcommand{\ya}[1]{\textcolor{blue}{\textbf{\underline{adiyoss}}: #1}\PackageWarning{TODO:}{#1!}}
\newcommand{\gab}[1]{\textcolor{green}{\textbf{\underline{gab}}: #1}\PackageWarning{TODO:}{#1!}}
}

\newcommand{\dem}{\textsc{Demucs}~}

\usepackage{bm}
\usepackage{amssymb,amsmath,graphicx,bbm,color,booktabs}
\usepackage{amsopn}

\usepackage{scalerel}
\DeclareMathOperator{\Conv}{Conv1d}

\DeclareMathOperator{\Convtr}{ConvTr1d}
\DeclareMathOperator{\ReLU}{Relu}
\DeclareMathOperator{\GLU}{GLU}
\DeclareMathOperator*{\bigadd}{\scalerel*{+}{\sum}}

\newcommand{\vn}{\bm{n}}

\newcommand{\vx}{\bm{x}}               
\newcommand{\vy}{\bm{y}}       \newcommand{\vyh}{\hat{\bm{y}}}        
\newcommand{\vz}{\bm{z}}       \newcommand{\vzh}{\hat{\bm{z}}}        

\renewcommand{\eqref}[1]{Eq.~(\ref{#1})}

\def \k{{\mathbf k}}

\title{Real Time Speech Enhancement in the Waveform Domain}
\name{Alexandre D\'efossez$^{1,2,3}$, Gabriel Synnaeve$^1$, Yossi Adi$^1$}
\address{
  $^1$Facebook AI Research~~$^2$INRIA~~$^3$PSL Research University}
\email{defossez,gab,adiyoss@fb.com}

\begin{document}

\maketitle

\begin{abstract}
We present a causal speech enhancement model working on the raw waveform that runs in real-time on a laptop CPU. The proposed model is based on an encoder-decoder architecture with skip-connections. It is optimized on both time and frequency domains, using multiple loss functions. Empirical evidence shows that it is capable of removing various kinds of background noise including stationary and non-stationary noises, as well as room reverb. Additionally, we suggest a set of data augmentation techniques applied directly on the raw waveform which further improve model performance and its generalization abilities. We perform evaluations on several standard benchmarks, both using objective metrics and human judgements. The proposed model matches state-of-the-art performance of both causal and non causal methods while working directly on the raw waveform.
\end{abstract}

\section{Introduction}
Speech enhancement is the task of maximizing the perceptual quality of speech signals, in particular by removing background noise. Most recorded conversational speech signal contains some form of noise that hinders intelligibility, such as street noise, dogs barking, keyboard typing, etc. Thus, speech enhancement is a particularly important task in itself for audio and video calls \cite{reddy2019scalable}, hearing aids \cite{reddy2017individualized}, and can also help automatic speech recognition (ASR) systems \cite{zorila2019investigation}. For many such applications, a key feature of a speech enhancement system is to run in real time and with as little lag as possible (online), on the communication device, preferably on commodity hardware. 

Decades of work in speech enhancement showed feasible solutions which estimated the noise model and used it to recover noise-deducted speech~\cite{lim1979enhancement,ephraim1984speech}. Although those approaches can generalize well across domains they still have trouble dealing with commons noises such as non-stationary noise or a babble noise which is encountered when a crowd of people are simultaneously talking. The presence of such noise types degrades hearing intelligibility of human speech greatly~\cite{krishnamurthy2009babble}. Recently, deep neural networks (DNN) based models perform significantly better on non-stationary noise and babble noise while generating higher quality speech in objective and subjective evaluations over traditional methods~\cite{pascual2017segan,phan2020improving}. Additionally, deep learning based methods have also shown to be superior over traditional methods for the related task of a single-channel source separation~\cite{luo2019conv,nachmani2020voice,demucs}.

Inspired by these recent advances, we proposed a real-time version of the \dem~\cite{demucs} architecture adapted for speech enhancement. It consists of a causal model, based on convolutions and LSTMs, with a frame size of 40ms, a stride of 16ms, and that runs faster than real-time on a single laptop CPU core. For audio quality purposes, our model goes from waveform to waveform, through hierarchical generation (using U-Net~\cite{unet} like skip-connections). We optimize the model to directly output the ``clean'' version of the speech signal while minimizing a regression loss function (L1 loss), complemented with a spectrogram domain loss~\cite{yamamoto2019parallel,yamamoto2019probability}. Moreover, we proposed a set of simple and effective data augmentation techniques: namely frequency band masking and signal reverberation. Although enforcing a vital real-time constraint on model run-time, our model yields comparable performance to state of the art model by objective and subjective measures.

Although, multiple metrics exist to measure speech enhancement systems these have shown to not correlate well with human judgements~\cite{reddy2019scalable}. Hence, we report results for both objective metrics as well as human evaluation. Additionally we conduct an ablation study over the loss and augmentation functions to better highlight the contribution of each part. Finally, we analyzed the artifacts of the enhancement process using Word Error Rates (WERs) produced by an Automatic Speech Recognition (ASR) model.

Results suggest that the proposed method is comparable to the current state-of-the-art model across all metrics while working directly on the raw waveform. Moreover, the enhanced samples are found to be beneficial for improving an ASR model under noisy conditions. 

\begin{figure*}
  \centering
  \def\pscale{0.7}
  \begin{subfigure}[b]{0.49\textwidth}
    \centering
    \scalebox{.8}{\begin{tikzpicture}[
    every node/.style={scale=\pscale},
    conv/.style={shape=trapezium,
        trapezium angle=70, draw, inner xsep=0pt, inner ysep=0pt,
        draw=black!90,fill=gray!5},
    deconv/.style={shape=trapezium,
        trapezium angle=-70, draw, inner xsep=0pt, inner ysep=0pt,
        draw=black!90,fill=gray!5},
    linear/.style={draw, inner xsep=1pt, inner ysep=1pt,
        draw=black!90,fill=green!5},
    rnn/.style={rounded corners=1pt,rectangle,draw=black!90,
        fill=blue!5,minimum width=0.6cm, minimum height=0.6cm},
    skip/.style={line width=0.2mm, ->},
]
    \def\yshift{0.3em}
    \def\base{9cm}
    \def\dec{0.45cm}

    \node (pad) at (0, -0.8cm) {};
    \node (base) at (0, 0) {};
    \def\sourcec{0.25 * cos(deg(0.3 * x)) * cos(deg(2 * x + 0.9 * cos(deg(4 * x)))}
    \begin{axis}[
        anchor=north,
        at=(base),
        scale=0.4,
        domain=-20:20,
        axis y line=none,
        axis x line=none,
        samples=200,
        color=black,
        height=2.5cm,
        width=\base + 8cm]
          \addplot[mark=none] {
              (
            rand
               )
         };
    \end{axis}
    \node (e1) [conv, minimum width=\base - \dec, anchor=south] at (0, 0)
        {$\mathrm{Encoder}_1(C_{\mathrm{in}}=1, C_{\mathrm{out}}=H)$};
    \node (e2) [conv, minimum width=\base - 2*\dec, anchor=south] at
        ($(e1.north) + (0, \yshift)$)
        {$\mathrm{Encoder}_2(C_{\mathrm{in}}=H, C_{\mathrm{out}}=2 H)$};
    \node (edots) [conv, minimum width=\base - 3*\dec, anchor=south] at
        ($(e2.north) + (0, \yshift)$)
        {$\ldots$};
    \node (e6) [conv, minimum width=\base - 4*\dec, anchor=south, inner ysep=2pt] at
        ($(edots.north) + (0, \yshift)$)
        {$\mathrm{Encoder}_L(C_{\mathrm{in}}=2^{L-2} H, C_{\mathrm{out}}=2^{L-1} H)$};

    \node (ls0) [rnn] at ($(e6.north) + (-0.33 * \base + 2 * \dec,0.4cm)$) {L};
    \foreach \k/\text in {1/S,2/T,3/M} {
        \tikzmath{
            int \prev;
            \prev=\k - 1;
        }
        \node (ls\k) [rnn,anchor=west] at ($(ls\prev.east) + (0.3cm, 0)$) {\text};
        \draw [->] (ls\prev) -- (ls\k);
    }
    \node (ls3) [anchor=west] at ($(ls3.east) + (0.1cm, 0)$) [align=left] {hidden size=$2^{L-1} H$\\2 layers};

    \node (d6) [deconv, minimum width=\base - 4*\dec, anchor=south, inner ysep=2pt] at
        ($(e6.north) + (0, 0.8cm)$) {$\mathrm{Decoder}_L(C_{\mathrm{in}}=2^{L-1} H, C_{\mathrm{out}}=2^{L-2} H)$};
    \node (ddots) [deconv, minimum width=\base - 3*\dec, anchor=south] at
        ($(d6.north) + (0, \yshift)$) {$\ldots$};
    \node (d2) [deconv, minimum width=\base - 2*\dec, anchor=south] at
        ($(ddots.north) + (0, \yshift)$) {$\mathrm{Decoder}_2(C_{\mathrm{in}}=2 H, C_{\mathrm{out}}=H)$};
    \node (d1) [deconv, minimum width=\base - \dec, anchor=south] at
        ($(d2.north) + (0, \yshift)$) {$\mathrm{Decoder}_1(C_{\mathrm{in}}=H, C_{\mathrm{out}}=1)$};

    % \draw [->, bend left]  (e1.west) -- (d1.west);
    \path[skip] (e1.west) edge[bend left=45] node [right] {} (d1.west);
    \path[skip] (e2.west) edge[bend left=45] node [right] {} (d2.west);
    \path[skip] (edots.west) edge[bend left=45] node [right] {} (ddots.west);
    \path[skip] (e6.west) edge[bend left=45] node [right] {} (d6.west);
    \newcommand\myoutput[3]{
        \begin{axis}[
            anchor=south,
            scale=0.4,
            at=#1,
            domain=-20:20,
            axis y line=none,
            axis x line=none,
            samples=200,
            height=2.5cm,
            color=#2,
            width=\base + 8cm]
            \addplot[mark=none] {
                #3
            };
        \end{axis}
    }
    \node (o3) at (d1.north) {};
    \myoutput{(o3)}{blue}{\sourcec}
\end{tikzpicture}}
    \caption{Causal \dem with the noisy speech as input on the bottom
    and the clean speech as output on the top. Arrows represents U-Net skip
    connections. $H$ controls the number of channels in the model and $L$ its depth.\label{fig:fullmodel}}
    
  \end{subfigure}%
  \hfill%
  \begin{subfigure}[b]{0.49\textwidth}
    \centering
    \scalebox{.8}{\begin{tikzpicture}[
    every node/.style={scale=\pscale},
    conv/.style={shape=trapezium,
        trapezium angle=70, draw, inner xsep=0pt, inner ysep=2pt,
        draw=black!90,fill=gray!5},
    deconv/.style={shape=trapezium,
        trapezium angle=-70, draw, inner xsep=0pt, inner ysep=0pt,
        draw=black!90,fill=gray!5},
    rewrite/.style={shape=rectangle,
        draw, inner xsep=11pt, inner ysep=3pt,
        draw=black!90,fill=gray!5},
    inout/.style={rounded corners=1pt,rectangle,draw=black!90,
        fill=violet!5,minimum width=0.6cm, minimum height=0.6cm},
    skip/.style={line width=0.2mm, ->},
    sum/.style      = {draw, circle, fill=gray!5, inner xsep=0pt, inner ysep=0pt},
]
    \def\yshift{0.3em}
    \def\base{8cm}
    \def\dec{0.55cm}
    \def\deltax{1cm}

    \node (base) at (0, 0cm) {};
    \node (pad) at (0cm, -1.4cm) {};
    \node (conv) [conv, minimum width=\base - 1.1 * \dec, anchor=south] at
      (base.north) {$\GLU(\Conv(C_{in}, 2 C_{in}, K=1, S=1))$};
    \node (deconv) [deconv, minimum width=\base - \dec, anchor=south] at
      ($(conv.north) + (0,\yshift)$) {$\Convtr(C_{in},C_{out}, K, S)$};

    \def\yshift{0.6em}
    \node (skip) [inout, anchor=north] at ($(conv.south) - (\deltax, 3 * \yshift)$) {$\mathrm{Encoder}_i$};
    \node (sum) [sum] at ($(conv.south) - (0, 1.5 * \yshift)$) {$\bigadd$};

    \draw[->]  (skip.north) -- ($(conv.south) - (\deltax, 1.5 * \yshift)$) -- (sum.west);

    \node (prev) [inout, anchor=north] at ($(conv.south) - (-\deltax, 3 * \yshift)$) {$\mathrm{Decoder}_{i+1}$ or LSTM};
    \draw[->]  (prev.north) -- ($(conv.south) - (-\deltax, 1.5 * \yshift)$) -- (sum.east);

    \draw[->]  (sum.north) -- (conv.south);

    \node (next) [inout, anchor=south] at ($(deconv.north) + (0, \yshift)$) {Output or ReLU then $\mathrm{Decoder}_{i-1}$};
    \draw[->]  (deconv.north) -- (next.south);
\end{tikzpicture}}
    \scalebox{.8}{\begin{tikzpicture}[
    every node/.style={scale=\pscale},
    conv/.style={shape=trapezium,
        trapezium angle=70, draw, inner xsep=0pt, inner ysep=0pt,
        draw=black!90,fill=gray!5},
    deconv/.style={shape=trapezium,
        trapezium angle=-70, draw, inner xsep=0pt, inner ysep=2pt,
        draw=black!90,fill=gray!5},
    rewrite/.style={shape=rectangle,
        draw, inner xsep=8pt, inner ysep=3pt,
        draw=black!90,fill=gray!5},
    inout/.style={rounded corners=1pt,rectangle,draw=black!90,
        fill=violet!5,minimum width=0.6cm, minimum height=0.6cm},
    skip/.style={line width=0.2mm, ->},
]
    \def\yshift{0.3em}
    \def\base{8cm}
    \def\deltax{1cm}
    \def\dec{0.55cm}

    \node (base) at (0, 0) {};
    \node (pad) at (0cm, -0.9cm) {};
    \node (conv) [conv, minimum width=\base - \dec, anchor=south] at
      (base.north) {$\ReLU(\Conv(C_{\mathrm{in}}, C_{\mathrm{out}}, K, S))$};
    \node (rewrite) [rewrite, minimum width=\base - 1.7 * \dec, anchor=south] at
      ($(conv.north) + (0,\yshift)$) {$\GLU(\Conv(C_{\mathrm{out}}, 2 C_{\mathrm{out}}, K = 1, S = 1))$};

    \def\yshift{0.6em}
    \node (skip) [inout, anchor=south] at ($(rewrite.north) + (-\deltax, \yshift)$) {$\mathrm{Decoder}_i$};
    \draw[->] ($(rewrite.north) + (-\deltax, 0)$) -- (skip.south);

    \node (prev) [inout, anchor=north] at ($(conv.south) - (0, \yshift)$) {Input or $\mathrm{Encoder}_{i-1}$};
    \draw[->]  (prev.north) -- (conv.south);

    \node (next) [inout, anchor=south] at ($(rewrite.north) + (\deltax, \yshift)$) {$\mathrm{Encoder}_{i+1}$ or LSTM};
    \draw[->]  ($(rewrite.north) + (\deltax, 0)$) -- (next.south);
\end{tikzpicture}}
    \caption{View of each encoder (bottom)
    and decoder layer (top). Arrows are connections
    to other parts of the model. 
    $C_{\mathrm{in}}$ (resp. $C_{\mathrm{out}}$) is the number of input channels (resp. output), $K$ the kernel size and $S$ the stride.\label{fig:modeldetail}}
  \end{subfigure}
  \caption{Causal \dem architecture on the left,
    with detailed representation of the encoder and decoder layers
  on the right. The on the fly resampling of the input/output by a factor of $U$ is not represented.
}
\end{figure*}
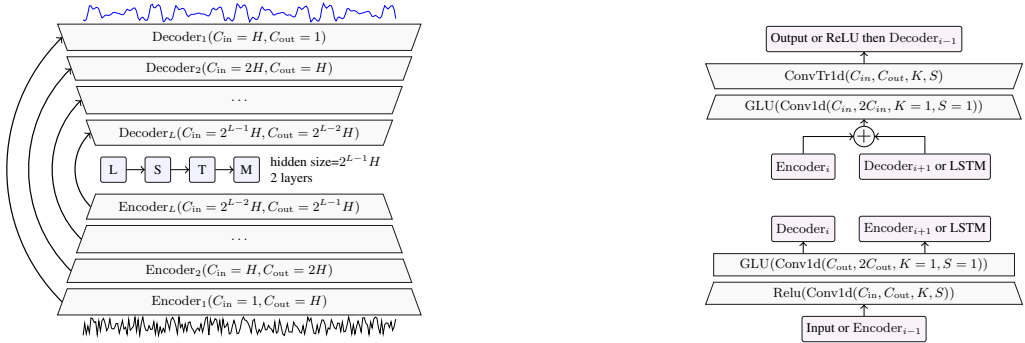

\vspace{-0.2cm}
\section{Model}
\vspace{-0.1cm}
\label{sec:model}

\subsection{Notations and problem settings}
We focus on monaural (single-microphone) speech enhancement that can operate in real-time applications. Specifically, given an audio signal $\vx\in\mathbb{R}^T$, composed of a clean speech $\vy\in\mathbb{R}^T$ that is corrupted by an additive background signal $\vn\in \mathbb{R}^T$ so that $\vx = \vy + \vn$. The length, T, is not a fixed value across samples, since the input utterances can have different durations. Our goal is to find an enhancement function $f$ such that $f(\vx) \approx \vy$. 

In this study we set $f$ to be the \dem architecture~\cite{demucs}, which was initially developed for music source separation, and adapt it to the task of causal speech enhancement, a visual description of the model can be seen in Figure~\ref{fig:fullmodel}. 

\vspace{-0.1cm}
\subsection{\dem architecture}
\vspace{-0.1cm}
\dem consists in a multi-layer convolutional encoder and decoder with U-net~\cite{unet} skip connections, and a sequence modeling network applied on the encoders' output. It is characterized by its number of layers $L$, initial number of hidden channels $H$, layer kernel size $K$ and stride $S$ and resampling factor $U$. The encoder and decoder layers are numbered from 1 to $L$ (in reverse order for the decoder, so layers at the same scale have the same index). As we focus on monophonic speech enhancement, the input and output of the model has a single channel only.

Formally, the encoder network $E$ gets as input the raw wave form and outputs a latent representation $E(\vx) = \vz$. 
Each layer consists in a convolution layer with a kernel size of $K$ and stride of $S$ with $2^{i-1} H$ output channels, followed by a ReLU activation, a ``1x1'' convolution with $2^{i} H$ output channels and finally a GLU~\cite{glu} activation that converts back the number of channels to $2^{i-1} H$, see Figure~\ref{fig:modeldetail} for a visual description. 

Next, a sequence modeling $R$ network takes the latent representation $\vz$ as input and outputs a non-linear transformation of the same size, $R(\vz) = LSTM(\vz) + \vz$, denoted as $\vzh$. The LSTM network consists of 2-layers and $2^{L-1} H$ hidden units. 
For causal prediction, we use an unidirectional LSTM, while for non causal models, we use
a bidirectional LSTM, followed by a linear layer to merge the both outputs.

Lastly, a decoder network $D$, takes as input $\vzh$ and outputs an estimation of clean signal $D(\vzh) = \vyh$. The i-th layer of the decoder takes as input $2^{i-1} H$ channels, and applies a 1x1 convolution
with $2^{i} H$ channels, followed by a GLU activation function that outputs $2^{i-1} H$ channels and finally
a transposed convolution with a kernel size of 8, stride of 4, and $2^{i-2} H$ output channels
accompanied by a ReLU function. For the last layer the output is a single channel and has no ReLU. A skip connection connects the output of the i-th layer of the encoder and the input of the i-th layer of the decoder, see Figure~\ref{fig:fullmodel}.

We initialize all model parameters using the scheme proposed by~\cite{kaiming_init}. 
Finally, we noticed that upsampling the audio by a factor $U$ before feeding it to the encoder improves accuracy.
We downsample the output of the model by the same amount. The resampling is done using a sinc interpolation filter~\cite{smith1984flexible}, as part of the end-to-end training, rather than a pre-processing step.

\vspace{-0.2cm}
\subsection{Objective}
\vspace{-0.2cm}
\label{sec:objective}
We use the L1 loss over the waveform together with a multi-resolution STFT loss over the spectrogram magnitudes similarly to the one proposed in~\cite{yamamoto2019parallel,yamamoto2019probability}. Formally, given $\vy$ and $\vyh$ be the clean signal and the enhanced signal respectively. We define the STFT loss to be the sum of the \emph{spectral convergence (sc)} loss and the \emph{magnitude} loss as follows,
\begin{equation}
    \label{eq:mrstft}
    \begin{aligned}
    &L_{\text{stft}}(\vy, \vyh) = L_{sc}(\vy, \vyh) + L_{mag}(\vy, \vyh) \\
    &L_{sc}(\vy, \vyh) = \frac{\| |\textit{STFT}(\vy)| - |\textit{STFT}(\vyh)|\|_F}{\||\textit{STFT}(\vy)| \|_F}\\
    &L_{mag}(\vy, \vyh) = \frac{1}{T}\| \log|\textit{STFT}(\vy)| - \log|\textit{STFT}(\vyh)|\|_1
    \end{aligned}
\end{equation}
where $\| \cdot \|_F$ and $\| \cdot \|_1$ are the Frobenius the $L_1$ norms respectively. We define the multi-resolution STFT loss to be the sum of all STFT loss functions using different STFT parameters. Overall we wish to minimize the following,
\begin{equation}
    \label{eq:obj}
    \frac{1}{T}[\|\vy - \vyh\|_1 + \sum_{i=1}^M L_{\text{stft}}^{(i)}(\vy, \vyh)]
\end{equation}
where $M$ is the number of STFT losses, and each $L_{\text{stft}}^{(i)}$ applies the STFT loss at different resolution with number of FFT bins $\in \{512, 1024, 2048\}$, hop sizes $\in \{50, 120,240 \}$, and lastly window lengths $\in \{240, 600, 1200\}$.
\vspace{-0.1cm}
\section{Experiments}

We performed several experiments to evaluate the proposed method against several highly competitive models. We report objective and subjective measures on the Valentini et al.~\cite{valentini2017noisy} and Deep Noise Suppression (DNS)~\cite{DNSChallenge2020} benchmarks. Moreover, we run an ablation study over the augmentation and loss functions. Finally, we assessed the usability of the enhanced samples to improve ASR performance under noisy conditions. Code and samples can be found in the following link: \textcolor{purple}{\url{https://github.com/facebookresearch/denoiser}}.

\setlength{\tabcolsep}{5.5pt}
\begin{table*}[t!]
	\centering
	\caption{Objective and Subjective measures of the proposed method against SOTA models using the Valentini benchmark~\cite{valentini2017noisy}.}
  \vspace{5pt}
  \label{tab:valent}
  \footnotesize
\begin{tabular}{l|ccccc|ccccc}
	\toprule
              							& PESQ & STOI (\%) & pred. & pred. & pred. & MOS & MOS & MOS  & Causal \\
              							& & & CSIG & CBAK & COVL & SIG & BAK & OVL & \\
	\midrule
Noisy             					    & 1.97 & 91.5  & 3.35 & 2.44 & 2.63 & 4.08 & 3.29 & 3.48 & -     \\
\midrule
SEGAN~\cite{pascual2017segan}     	    & 2.16 & -     & 3.48 & 2.94 & 2.80 & - & - & - &  No     \\
Wave U-Net~\cite{macartney2018improved} & 2.40  & -     & 3.52 & 3.24 & 2.96 & - & - & - & No     \\
SEGAN-D~\cite{phan2020improving}		& 2.39 & -     & 3.46 & 3.11 & 3.50 & - & - & - & No     \\
MMSE-GAN~\cite{soni2018time}  			& 2.53 & 93    & 3.80 & 3.12 & 3.14 & - & - & - & No     \\
MetricGAN~\cite{fu2019metricgan} 		& 2.86 & -     & 3.99 & 3.18 & 3.42 & - & - & - & No     \\
DeepMMSE~\cite{zhang2020deepmmse}       & 2.95 & 94  & \textbf{4.28} & \textbf{3.46} & \textbf{3.64} & - & - & -  & No \\
\dem ($H{=}64$, $S{=}2$ ,$U{=}2$)    & \textbf{3.07}  & \textbf{95} & \textbf{4.31} & 3.4 & \textbf{3.63} & 4.02  & 3.55 & 3.63 & No     \\
\midrule
Wiener            					& 2.22 & 93    & 3.23 & 2.68 & 2.67 & - & - & - & Yes \\
DeepMMSE~\cite{zhang2020deepmmse}   & 2.77 & 93  & 4.14 & \textbf{3.32} & \textbf{3.46} & 4.11 & \textbf{3.69} & \textbf{3.67} & Yes \\
\dem ($H{=}48$,$S{=}4$, $U{=}4$)    &  \textbf{2.93}    & \textbf{95}  & \textbf{4.22} & 3.25 & \textbf{3.52} & 4.08 & 3.59 & 3.40 & Yes    \\
\dem ($H{=}64$,$S{=}4$, $U{=}4$)    &  2.91    & \textbf{95}  & \textbf{4.20} & 3.26 & \textbf{3.51} & 4.03 & \textbf{3.69} & 3.39 & Yes    \\
\dem ($H{=}64$,$S{=}4$, $U{=}4$) + dry=0.05    &  2.88    & \textbf{95} & 4.14 & 3.21 & \textbf{3.54} & 4.10 & 3.58 & \textbf{3.72} & Yes    \\
\dem ($H{=}64$,$S{=}4$, $U{=}4$) + dry=0.1    &  2.81    & \textbf{95}  & 4.07 & 3.10 & 3.42 & \textbf{4.18} & 3.45 & 3.60 & Yes    \\
\bottomrule
\end{tabular}
\end{table*}

\subsection{Implementation details}
\label{sec:impl_details}
\paragraph*{Evaluation Methods}
We evaluate the quality of the enhanced speech using both objective and subjective measures. For the objective measures we use: (i) PESQ: Perceptual evaluation of speech quality, using the wide-band version recommended in ITU-T P.862.2~\cite{recommendation2001perceptual} (from –0.5 to 4.5) (ii) Short-Time Objective Intelligibility (STOI)~\cite{taal2011algorithm} (from 0 to 100) (iii) CSIG: Mean opinion score (MOS) prediction of the signal distortion attending only to the speech signal~\cite{hu2007evaluation} (from 1 to 5). (iv) CBAK: MOS prediction of the intrusiveness of background noise~\cite{hu2007evaluation} (from 1 to 5). (v) COVL: MOS prediction of the overall effect~\cite{hu2007evaluation} (from 1 to 5). 

For the subjective measure, we conducted a MOS study as recommended in ITU-T P.835~\cite{recommendation2003subjective}. For that, we launched a crowd source evaluation 
using the CrowdMOS package~\cite{protasioribeiro2011crowdmos}. We randomly sample 100 utterances and each one was scored by 15 different raters along three axis: level of distortion, intrusiveness of background noise, and overall quality. Averaging results across all annotators and queries gives the final scores. 

\noindent{\bf Training\quad}
We train the \dem model for 400 epochs on the Valentini~\cite{valentini2017noisy} dataset and 250 on the DNS~\cite{DNSChallenge2020} dataset. 
We use the L1 loss between the predicted and ground truth clean speech waveforms, and for the Valentini dataset, also add the STFT loss described in Section~\ref{sec:objective} with a weight of $0.5$. We use the Adam optimizer with a step size of $3\mathrm{e}{-}4$, a momentum of $\beta_1=0.9$ and a denominator momentum $\beta_2=0.999$. For the Valentini dataset, we use the original validation set and keep the best model, for the DNS dataset we train without a validation set and keep the last model. The audio is sampled at 16 kHz.

\noindent{\bf Model\quad}
We use three variants of the \dem architecture described in Section~\ref{sec:model}. For the non causal \dem, we take $U{=}2$, $S{=}2$, $K{=}8$, $L{=}5$ and $H{=}64$. For the causal \dem, we take $U{=}4$, $S{=}4$, $K{=}8$ and $L{=}5$, and either $H{=}48$, or $H{=}64$.   We normalize the input by its standard deviation before feeding it to the model and scale back the output by the same factor. For the evaluation of causal models, we use an online estimate of the standard deviation.
With this setup, the causal \dem processes audio has a frame size of 37 ms and a stride of 16 ms.

\noindent{\bf Data augmentation\quad}
We always apply a random shift between 0 and $S$ seconds. 
The \emph{Remix} augmentation shuffles the noises within one batch to form new noisy mixtures.
\emph{BandMask} is a band-stop filter with a stop band between $f_0$ and $f_1$, sampled to remove 20\% of the frequencies uniformly in the mel scale.
This is equivalent, in the waveform domain, to the SpecAug augmentation~\cite{specaug} used for ASR training.
\emph{Revecho}: 
given an initial gain $\lambda$, early delay $\tau$ and RT60, it adds to the noisy signal a series of $N$ decaying echos of the clean speech and noise. The $n$-th echo has a delay of $n \tau + \mathrm{jitter}$ and a gain of $\rho^n\lambda$. $N$ and $\rho$ are chosen so that when the total delay reaches RT60, we have $\rho^N \leq 1\mathrm{e}{-}3$. $\lambda$, $\tau$ and RT60 are sampled uniformly respectively over $[0, 0.3]$, $[10, 30]$ ms, $[0.3, 1.3]$ sec.

We use the random shift for all datasets, Remix and Banmask for Valentini~\cite{valentini2017noisy}, and Revecho only for DNS~\cite{DNSChallenge2020}.

\noindent{\bf Causal streaming evaluation\quad}
In order to test our causal model in real conditions we use a specific streaming implementation
at test time. Instead of normalizing by the standard deviation of the audio, we use the
standard deviation up to the current position (i.e. we use the cumulative standard deviation).
We keep a small buffer of past input/output to limit the side effect of the sinc resampling filters. 
For the input upsampling, we also use a 3ms lookahead, which takes the total frame size of the model to 40 ms. When applying the model to a given frame of the signal, the rightmost part of the output is invalid, because future audio is required to compute properly the output of the transposed convolutions. Nonetheless, we noticed that using this invalid part as a padding for the streaming downsampling greatly improves the PESQ. The streaming implementation is pure PyTorch. Due to the overlap between frames, care was taken to cache the output of the different layers as needed.

\vspace{-0.2cm}
\subsection{Results}
\vspace{-0.1cm}
Table~\ref{tab:valent} summarizes the results for Valentini~\cite{valentini2017noisy} dataset using both causal and non-causal models. Results suggest that \dem matched current SOTA model (DeepMMSE~\cite{zhang2020deepmmse}) using both objective and subjective measures, while working directly on the raw waveform and without 
using extra training data. Additionally, \dem is superior to the other baselines methods, (may they be causal or non-causal), by a significant margin. We also introduce a dry/wet knob, i.e. we output
$\mathrm{dry} \cdot x + (1 - \mathrm{dry})\cdot \hat{y}$, which allows to control the trade-off between
noise removal and conservation of the signal. We notice that a small amount of bleeding (5\%) improves the overall perceived quality.

\begin{table}[t!]
	\small
	\centering
	\caption{Subjective measures of the proposed method with different treatment of reverb, on the DNS blind test set~\cite{DNSChallenge2020}.
	Recordings are divided in 3 categories: no reverb, reverb (artificial)
	and real recordings. We report the OVL MOS.
	All models are causal. For \dem, we take $U{=}4$, $H{=}64$ and $S{=}4$.}
  \vspace{2pt}
  \label{tab:dns}
\resizebox{\columnwidth}{!}{%
\begin{tabular}{lccc}
\toprule
                                            & No Rev. & Reverb & Real Rec. \\
\midrule
Noisy                    					& 3.1 & 3.2 &  2.6 \\
NS-Net~\cite{xia2020weighted,DNSChallenge2020}      	& 3.2 & 3.1   & 2.8 \\
\midrule
\dem, no reverb     					&  \textbf{3.7} & 2.7   &  3.3   \\
\dem, remove reverb     					&  3.6 & 2.6   &  3.2   \\
\dem, keep reverb     					&  3.6 & 3.0   &  3.2   \\
\dem, keep 10\% rev.     					&  3.3 & \textbf{3.3}   &  3.1   \\
\dem, keep 10\% rev., two sources    		& 3.6 & 2.8   &  \textbf{3.5}   \\
\bottomrule
\end{tabular}
}
\vspace{-0.4cm}
\end{table}

We present on Table~\ref{tab:dns} the overall MOS evaluations on
the 3 categories of the DNS~\cite{DNSChallenge2020} blind test set: no reverb (synthetic mixture without reverb), reverb (synthetic mixture, with
artificial reverb) and real recordings. We test different strategies
for the reverb-like Revecho augmentation described in Section~\ref{sec:impl_details}.
We either ask the model to remove it (dereverberation), keep it, or keep only part of it. Finally, we either
add the same reverb to the speech and noise or use different jitters
to simulate having two distinct sources.
We entered the challenge with the ``remove reverb'' model,
with poor performance on the Reverb category due to dereverberation
artifacts\footnote{MOS for the baselines differ from the challenge
as we evaluated on 100 examples per category, and used
ITU-T P.835 instead of P.808.}.
Doing partial dereverberation improves
the overall rating, but not for real recordings (which have typically
less reverb already). On real recordings, simulating reverb with two sources improves the ratings.

\vspace{-0.1cm}
\subsection{Ablation}
In order to better understand the influence of different components in the proposed model on the overall performance, we conducted an ablation study over the augmentation functions and loss functions. We use the causal \dem version and report PESQ and STOI for each of the methods. Results are presented in Table~\ref{tab:ablation}. Results suggest that each of the components contribute to overall performance, with the STFT loss and time shift augmentation producing the biggest increase in performance. Notice, surprisingly the \emph{remix} augmentation function has a minor contribution to the overall performance. 

\begin{table}[t!]
	\small
	\centering
	\caption{Ablation study for the causal \dem  model with $H{=}64, S{=}4, U{=}4$ using the Valentini benchmark~\cite{valentini2017noisy}.}
  \vspace{2pt}
  \label{tab:ablation}
\resizebox{\columnwidth}{!}{%
\begin{tabular}{lcc}
\toprule
											&  PESQ & STOI ($\%$)\\
\midrule
Reference     					&  2.91   & 95   \\
no BandMask (BM)     					&  2.87   & 95   \\
no BM, no remix    			&  2.86   & 95   \\
no BM, no remix,  no STFT loss    			&  2.68   & 94   \\
no BM, no remix, no STFT loss , no shift    			&  2.38   & 93   \\
\bottomrule
\end{tabular}
}
\end{table}

\vspace{-0.1cm}
\subsection{Real-Time Evaluation}
\label{sec:rte}
We computed the Real-Time Factor (RTF, e.g. time to enhance a frame divided by the stride) under the streaming setting to better match real-world conditions. We benchmark this implementation on a quad-core Intel i5 CPU (2.0 GHz, up to AVX2 instruction set). The RTF is 1.05 for the $H{=}64$ version, while for the $H{=}48$ the RTF is 0.6. When restricting execution to a single core, the $H{=}48$ model still achieves a RTF of 0.8, making it realistic to use in real conditions, for instance along a video call software. We do not provide RTF results for DeepMMSE~\cite{zhang2020deepmmse} since no streaming implementation was provided by the authors, thus making it an unfair comparison.

\vspace{-0.1cm}
\subsection{The effect on ASR models}
Lastly, we evaluated the usability of the enhanced samples to improve ASR performance under noisy conditions. To that end, we synthetically generated noisy data using the \textsc{Librispeech} dataset~\cite{panayotov2015librispeech} together with noises from the test set of the DNS~\cite{DNSChallenge2020} benchmark. We created noisy samples in a controlled setting where we mixed the clean and noise files with SNR levels $\in \{0, 10, 20, 30\}$. For the ASR model we used a Convolutions and Transformer based acoustic model which get states-of-the-art results on \textsc{Librispeech}, as described in~\cite{synnaeve2019end}. To get Word Error Rates (WERs) we follow a simple Viterbi (argmax) decoding with neither language model decoding nor beam-search. That way, we can better understand the impact of the enhanced samples on the acoustic model. Results are depicted in Table~\ref{tab:asr}. \dem is able to recover up to 51\% of the WER lost to the added noise at SNR 0, and recovering in average 41\% of the WER on dev-clean and 31\% on dev-other. As the acoustic model was not retrained on denoised data, those results show the direct applicability of speech enhancement to ASR systems as a black-box audio preprocessing step.

\begin{table}[t!]
  \small
  \centering
  \caption{ASR results with a state-of-the-art acoustic model, Word Error Rates without decoding, no language model. Results on the \textsc{LibriSpeech} validation sets with added noise from the test set of DNS, and enhanced by \dem. \label{tab:asr}}
  \vspace{2pt}
\resizebox{\columnwidth}{!}{%
\begin{tabular}{lrrrr}
\toprule
Viterbi WER on & dev-clean & enhanced & dev-other & enhanced \\
\midrule
original (no noise) & 2.1 & 2.2 & 4.6 & 4.7 \\
noisy SNR 0 & 12.0 & 6.9 & 21.1 & 14.7 \\
noisy SNR 10 & 9.8 & 6.3 & 18.4 & 13.1 \\
noisy SNR 20 & 5.2 & 4.0 & 11.7 & 9.4 \\
noisy SNR 30 & 3.3 & 2.9 & 7.6 & 7.2 \\
\bottomrule
\end{tabular}
}
\vspace{-0.4cm}
\end{table}
\vspace{-0.1cm}
\section{Related Work}
\vspace{-0.1cm}

Traditionally speech enhancement methods generate either an enhanced version of the magnitude spectrum or produce an estimate of the ideal binary mask (IBM) that is then used to enhance the magnitude spectrum~\cite{ephraim1984speech,hu2006subjective}.

Over the last years, there has been a growing interest towards DNN based methods for speech enhancement~\cite{wang2015deep, weninger2015speech, xu2014regression, rethage2018wavenet, macartney2018improved, nicolson2019deep, germain2018speech, fu2019metricgan, nikzad2020deep, pascual2017segan, wang2018investigating, phan2020improving, baby2019sergan, soni2018time, rethage2018wavenet}. In~\cite{wang2015deep} a deep feed-forward neural network was used to generate a frequency-domain binary mask using a cost function in the waveform domain. Authors in~\cite{xu2017multi} suggested to use a multi-objective loss function to further improve speech quality. Alternatively authors in~\cite{weninger2014discriminatively, weninger2015speech} use a recursive neural network (RNN) for speech enhancement. In~\cite{pascual2017segan} the authors proposed an end-to-end method , namely Speech Enhancement Generative Adversarial Networks (SEGAN) to perform enhancement directly from the raw waveform. The authors in~\cite{wang2018investigating, phan2020improving, baby2019sergan, soni2018time} further improve such optimization. In~\cite{rethage2018wavenet} the authors suggest to use a WaveNet~\cite{oord2016wavenet} model to perform speech denoising by learning a function to map noisy to clean signals.

While considering causal methods, the authors in~\cite{tan2018convolutional} propose a convolutional recurrent network at the spectral level for real-time speech enhancement, while Xia, Yangyang, et al.~\cite{xia2020weighted} suggest to remove the convolutional layers and apply a weighted loss function to further improve results in the real-time setup. Recently, the authors in~\cite{zhang2020deepmmse} provide impressive results for both causal and non-causal models using a minimum mean-square error noise power spectral density tracker, which employs a temporal convolutional network (TCN) a priori SNR
estimator.

\vspace{-0.1cm}
\section{Discussion}

We have showed how \dem, a state-of-the-art architecture developed for music source
separation in the waveform domain, could be turned into a causal speech enhancer, processing
audio in real time on consumer level CPU. We tested \dem on the standard Valentini benchmark and achieved state-of-the-art result without using extra training data. We also test our model in real reverberant conditions with the DNS dataset. We empirically demonstrated how augmentation techniques (reverb with two sources, partial dereverberation) can produce a significant improvement in subjective evaluations. Finally, we showed that our model can improve the performance of an ASR model in noisy conditions even without retraining of the model. 

\clearpage
\bibliographystyle{IEEEtran}
\bibliography{mybib}

\end{document}